%% file: main.tex
\documentclass[conference]{IEEEtran}
\IEEEoverridecommandlockouts
\usepackage{cite}
\usepackage{amsmath,amssymb,amsfonts}
\usepackage{graphicx}
\usepackage{textcomp}
\usepackage{xcolor}
\usepackage{algorithm}
\usepackage{algpseudocode}
\usepackage{amsmath}
\usepackage{lipsum}
\usepackage{acronym}

\def\BibTeX{{\rm B\kern-.05em{\sc i\kern-.025em b}\kern-.08em
    T\kern-.1667em\lower.7ex\hbox{E}\kern-.125emX}}
\begin{document}
\input{acronyms}

\title{Hardware architecture and routing-aware training for optimal memory usage: a case study\\

\thanks{This work is supported by SNSF Starting Grant Project UNITE (TMSGI2-211461)}
}

\author{\IEEEauthorblockN{
Jimmy Weber\IEEEauthorrefmark{1}\IEEEauthorrefmark{3},
Theo Ballet\IEEEauthorrefmark{1}\IEEEauthorrefmark{2}\IEEEauthorrefmark{3}, 
Melika Payvand\IEEEauthorrefmark{1}}

\IEEEauthorblockA{
\IEEEauthorrefmark{1}Institute of Neuroinformatics,
University of Zurich and ETH Zurich, Zurich, Switzerland  \\
\IEEEauthorrefmark{2}University of Paris Saclay, Paris, France\\
\IEEEauthorrefmark{3} These authors have contributed equally to this work. \\
Email: \{jimmy, melika\}@ini.uzh.ch }}

\maketitle

\begin{abstract}

Efficient deployment of neural networks on resource-constrained hardware demands optimal use of on-chip memory. In event-based processors, this is particularly critical for routing architectures, where substantial memory is dedicated to managing network connectivity. While prior work has focused on optimizing event routing during hardware design, optimizing memory utilization for routing during network training remains underexplored. Key challenges include: (i) integrating routing into the loss function, which often introduces non-differentiability, and (ii) computational expense in evaluating network mappability to hardware.
We propose a hardware-algorithm co-design approach to train routing-aware neural networks. To address challenge (i), we extend the \emph{DeepR} training algorithm, leveraging dynamic pruning and random re-assignment to optimize memory use. For challenge (ii), we introduce a proxy-based approximation of the mapping function to incorporate placement and routing constraints efficiently.
We demonstrate our approach by optimizing a network for the Spiking Heidelberg Digits (SHD) dataset using a small-world connectivity-based hardware architecture as a case study. The resulting network, trained with our routing-aware methodology, is fully mappable to the hardware, achieving 5\% more accuracy using the same number of parameters, and iso-accuracy with 10x less memory usage, compared to non-routing-aware training methods.
This work highlights the critical role of co-optimizing algorithms and hardware to enable efficient and scalable solutions for constrained environments.

\end{abstract}


\section{Introduction}

Efficient computation relies on the optimal utilization of the resources provided by the underlying hardware substrate. This becomes more critical in constrained environments where memory and energy budgets are limited. Specifically, when deploying a neural network on a hardware substrate with such constraints, it is essential to carefully plan how the network will be mapped onto it. Optimizing resource allocation, particularly memory and power, is crucial to enabling the deployment of larger networks on hardware with fixed capacities.
We study this problem on the example of neuromorphic event-based processors~\cite{Davies_etal18_loihi,Schemmel_etal_2010_brainscales,basu2022spiking,moradi_etal2017_dynapse,painkras_etal2013_spinnaker,Merolla_etal14_truenorth}, but the resulting findings are more broadly applicable. 
In these processors, a key computing unit is the neuron that integrate the incoming information and generate an output event (or spike) once the integration value passes a certain threshold (aka \ac{LIF} model). 
Typically, groups of these Neurons are organized into ``cores'', and multiple cores are combined in a multi-core architecture to enable modular scalability~\cite{Davies_etal18_loihi,Schemmel_etal_2010_brainscales,benjamin_etal14_neurogrid,Furber14_spinnaker,moradi_etal2017_dynapse}. 
Communication between neurons across these systems is facilitated by a routing mechanism, commonly implemented through a Network-on-Chip (NoC)~\cite{park_etal2016_hiAER}. 
Consequently, a significant memory footprint of such hardware is allocated to managing network connectivity and storing parameter values, such as neural network weights. 
Different event-based NoC schemes have been developed to optimize various metrics such as latency of communication, flexibility of programming, and memory and power footprint~\cite{park_etal2016_hiAER,moradi_etal2017_dynapse,dalgaty_etal_2024_mosaic,Davies_etal18_loihi}.
As the memory to store the network connectivity amounts for a large portion of the on-chip memory (e.g., 32\% on DYNAP-SE~\cite{moradi_etal2017_dynapse}), optimizing for memory footprint at the time of the design is important and has been the focus of recent work~\cite{leite_etal_2022_cortical,zhesu_etal2024_multicast}. 

However, it is also important to do such optimization from the algorithmic stand point.
Specifically, given memory blocks with unique placement, i.e. hardware architecture with certain routing, how can the network be optimized to ensure hardware mappability with the optimal use of memory resources.
One promising approach is to include hardware-specific routing constraints into the loss function, such that stochastic gradient-descent can minimize the routing mismatch between neural network model and hardware architecture.
However, the mapping function that accounts for routing and placement is often non-differentiable, posing a significant challenge for its optimization. 

One solution that does not require differentiability is evolutionary strategies~\cite{schuman_etal2020_eons,salimans_etal_2017_ES}, but such methods tend to be computationally more expensive and slower than gradient-based methods. To address this limitation, the \emph{DeepR} algorithm~\cite{bellec_etal_2017_deepR} was introduced to optimize neural networks while considering the memory footprint, i.e., the number of available memory units. 

Here, we take a step further and re-purpose the \emph{DeepR} algorithm, to not only take into account the number of memory elements while optimizing the network, but also the feasibility of mapping the network on the hardware architecture, with certain memory placement. 
To simplify the mapping function evaluation which is computationally intensive, we define a proxy to the mapping function.
We apply the augmented \emph{DeepR} with the proxy function to the case study of a hardware architecture, based on the small-world connectivity (locally-dense and globally-sparse) , i.e. the Mosaic architecture~\cite{dalgaty_etal_2024_mosaic}. To evaluate our method, we optimize a fully routing-aware small-world network to classify the spoken digits of the \ac{SHD} dataset.
We show that (i) the recurrent network trained with our novel method becomes fully mappable to the hardware architecture, 
and (ii) compared to a non-routing optimized network, for the same memory footprint it achieves 5\% more accuracy point on \ac{SHD}, and for the same accuracy, it requires about one order of magnitude less memory elements.

\section{Background}
\subsection*{Neuromorphic Memory Mosaic Architecture}

\begin{figure}
\centering
\includegraphics[width=\columnwidth]{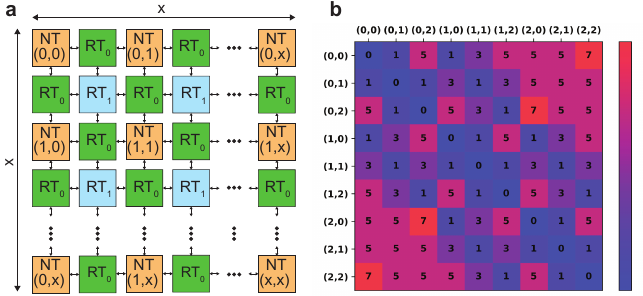}
\caption{Mosaic hardware architecture as our case study, with a small-world connectivity layout. a) Details of the Mosaic architecture with distributed Neuron Tiles (NT) and Routing Tiles ($RT_0$ and $RT_1$). Each NT integrates incoming messages from its neighbors and send its output to the fabricate through its routing neighbors ($RT_0$). $RT_1$s interface with $RT_0$s to pass along the spikes to other RTs. b) The cost of communication between NTs (number of hops required to go from one to another) at different locations for a Mosaic architecture of size 3 $\times$ 3. }
\label{fig:fig1_mosaic}
\end{figure}

The \emph{Mosaic} architecture is a two-dimensional systolic array of modular computing cores (\acp{NT}), alternated with routers (\acp{RT}), which connect them together (as illustrated in Fig.~\ref{fig:fig1_mosaic}a)~\cite{dalgaty_etal_2024_mosaic}. 
The \acp{NT} (shown in orange squares) host a cluster of all-to-all connected spiking neurons arranged in a crossbar array, implementing the \ac{LIF} model, communicating to the neighboring routers. The \acp{RT} (shown in blue and green squares) are also crossbar arrays, either forwarding or blocking the incoming spikes, directing them into the computing fabric. The Mosaic therefore takes advantage of locality, both in computing and communication, to reduce power usage while effectively implementing a small-world-like densely-local and sparsely global connectivity.

\subsection*{DeepR}
\emph{DeepR}~\cite{bellec_etal_2017_deepR} is a method developed to train neural networks under strict memory constraints, by enforcing a fixed limit on the total number of active connections in the network. We look at this constraint as a connection sparsity, defined as the ratio of the number of active connections to the total possible ones, $S(\theta) = \frac{|\theta|_0}{N^2}$, where \( \theta \in \mathbb{R}^{N \times N} \) represents the network parameters for a network with \( N \) neurons, \(|\theta|_0\) is the \(\text{L}_0\)-norm of \(\theta\), counting its number of non-zero elements, and \( S(\theta) \) computes the sparsity of the weight matrix.

\emph{DeepR} adjusts the network connectivity during training, while ensuring the network operates within a pre-defined memory budget (\(\text{L}_0\)-norm of \(\theta\)) or target sparsity \( \hat{s} \). The following pseudo-code outlines the iterative steps of the \emph{DeepR} algorithm.

\begin{algorithm}[H]
\caption{Simplified DeepR (adapted for connection sparsity)} 
\begin{algorithmic}[1]
\State Initialize network weights \( \theta \) such that \( S(\theta) = \hat{s} \)
\While{training not converged}
    \State Perform SGD on \( \mathcal{L} + \lambda \cdot |\theta|_1 \)
    \State Prune weights: \( \theta' \gets f(\theta) \) s.t. \( S(\theta') \leq \hat{s} \)
    \State Reassign connections: \( \theta'' \gets g(\theta') \) s.t. \( S(\theta'') = \hat{s} \)
    \State Update weights: \( \theta \gets \theta'' \)
\EndWhile
\end{algorithmic}
\end{algorithm}

The \emph{DeepR} algorithm begins by initializing the network weights \( \theta \) such that the sparsity constraint \( S(\theta) = \hat{s} \) is satisfied (Line 1). During each epoch, stochastic gradient descent (SGD) is performed on a modified loss function \( \mathcal{L} + \lambda \cdot |\theta|_1 \), where the term \( \lambda \cdot |\theta|_1 \) encourages sparsity (Line 3). The pruning function \( f(\cdot) \) removes connections with weights below a fixed threshold (Line 4), and the reassignment function \( g(\cdot) \) reallocates the inactive connections, ensuring the target sparsity \( S(\theta'') = \hat{s} \) (Lines 5). Finally, the updated weights \( \theta'' \) replace the previous weights \( \theta \) to complete the epoch (Line 6).

\section{Methods}
\subsection*{Routing in Mosaic}
\begin{figure*}[h]
\centering
\includegraphics[width = \textwidth]{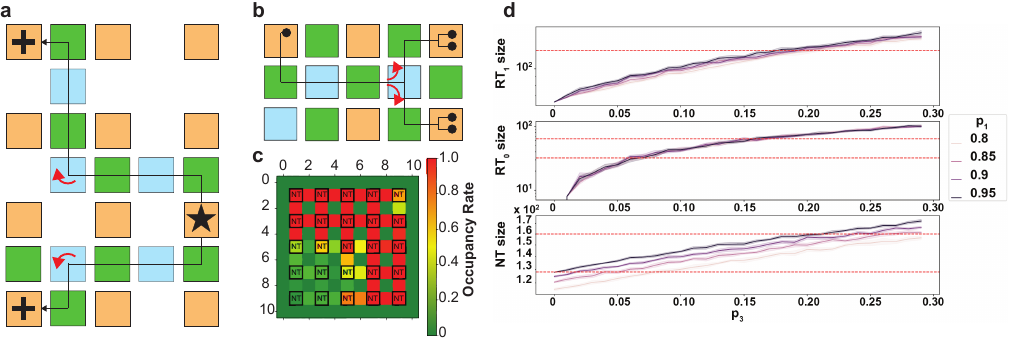}
\caption{Routing information on the Mosaic architecture. a) We use the 1-turn algorithm for routing spikes, where the routing path from the source to destination only takes one turn (red arrow). We use $RT_1$s (blue squares) for turning, and keep $RT_0$s (green squares) as no-turn routers. b) We use a shared-path routing, such that if two destinations have an overlapping path, the path is shared for the longest possible distance, before taking a turn to the two destinations. c) The occupancy rate of each tile for an example mapped network onto the Mosaic architecture. The routing algorithm calculates the required resource per NT and RT, and checks whether the network is mappable on the hardware. d) The minimum number of required memory resources for the NT and RT as a function of the sparsity of connections for each hop, to ensure mappability of the network on the Mosaic hardware. RT and NT sizes refer to input size of a square crossbar array.}
\label{fig:fig2_routing}
\end{figure*}

To communicate information from a neuron in \( \text{NT}_i \) to a set of neurons in \( \text{NT}_j \), the information is routed through a series of \acp{RT}. The number of RTs that the information traverses is referred to as \textit{Hops}, denoted by \( H(\text{NT}_i, \text{NT}_j) \). Fig.~\ref{fig:fig1_mosaic}b indicates the hop distance between a few neurons tiles of different coordinates.
Note that the memory required to route the activity between \acp{NT} scales linearly with the number of hops.

The routing algorithm used in the Mosaic architecture is a 1-turn algorithm inspired by ~\cite{seo_etal_2005_1turn}. 
Through the \acp{RT}, information from the source \ac{NT} is routed first along the \(x\)-coordinate of the destination \ac{NT}, and then along its \(y\)-coordinate, as illustrated in Fig.~\ref{fig:fig2_routing}a. 
This stepwise routing ensures efficient communication, while minimizing resource usage. Two types of \acp{RT} are employed in the Mosaic architecture:
\begin{itemize}
    \item \textcolor{green!50!black}{$\textbf{RT}_0$}: These tiles are adjacent to \acp{NT} and can communicate directly with them. Additionally, $\text{RT}_0$ tiles can transmit information either vertically or horizontally to other \acp{RT}, but do not support turning operations.
    \item \textcolor{cyan}{$\textbf{RT}_1$}: These tiles are designed exclusively for transmitting activities to other \acp{RT}. They can transmit data both in the horizontal or vertical directions, and are capable of performing turns.
\end{itemize}

To further optimize memory usage, Mosaic employs shared-path routing. When information is transmitted from a neuron to multiple destination neurons, whether within the same or to a different \ac{NT}, the routing algorithm reuses the same path for the longest possible distance before it diverges to the different destinations (see Fig. \ref{fig:fig2_routing}b).

Given a network connectivity matrix, the routing algorithm maps the network onto the Mosaic architecture, and checks whether the routing constraints are met (see Fig. \ref{fig:fig2_routing}c).

\subsection*{Approximating the Routing constraints}

We define the sparsity of connections between neurons in different \acp{NT} with a hop distance of \( d \) as:
\[p_d(\theta) = S(\theta_{i,j}) \big|_{H(\text{NT}_i, \text{NT}_j) = d}\]
where \( \theta_{i,j} \) is the connectivity between the neurons located in \( \text{NT}_i \) and \( \text{NT}_j \). 

Figure~\ref{fig:fig2_routing}d illustrates the required memory resource for \acp{NT} and \acp{RT} to ensure the feasibility of routing for different sets of \emph{sparsity profile}  \( P(\theta) = \{p_d(\theta) \mid d \in \{0, \ldots,  d_{\text{max}}\}\} \). The analysis demonstrates that the total resources required for routing can be accurately approximated by the values of \( P \), as the standard deviation of memory resource usage across networks with identical \( P \) values is minimal.

This result indicates that \( P \) effectively captures the routing constraints, serving as a reliable and computationally efficient proxy for estimating resource needs in Mosaic networks.

\subsection*{DeepR adaptation}
For a given Mosaic architecture with certain number of memory elements per \ac{NT} and \ac{RT}, its constraints can be approximated by a target sparsity profile \( \hat{P} \). To ensure compliance with these constraints during training, we extend the \emph{DeepR} algorithm by incorporating \( \hat{P} \) as a hardware-aware sparsity constraint (Algorithm 2).

\begin{algorithm}[H]
\caption{DeepR \textcolor{blue}{(HW-architecture-aware adaptation)}}
\begin{algorithmic}[1]
\State Initialize network weights \( \theta \) such that \textcolor{blue}{\( P(\theta) = \hat{P} \)}
\While{training not converged}
    \State Perform SGD on \( \mathcal{L} + \lambda \cdot |\theta|_1 \)
    \State Prune weights: \( \theta' \gets f(\theta) \) s.t. \textcolor{blue}{\( P(\theta') \leq \hat{P} \)}
    \State Reassign connections: \( \theta'' \gets g(\theta') \) s.t. \textcolor{blue}{\( P(\theta'') = \hat{P} \)}
    \State Update weights: \( \theta \gets \theta'' \)
\EndWhile
\end{algorithmic}
\end{algorithm}
The conditions \( P(\theta) = \hat{P} \) and \( P(\theta) \leq \hat{P} \) define element-wise relationships between the sparsity constraints across all hop distances, specifically:
\[
P(\theta) \leq \hat{P} \iff p_d(\theta) \leq \hat{p}_d, \, \forall \, d \in \{0, \ldots,  d_{\text{max}}\}
\]

\section{Results}

To evaluate our routing-aware optimization method, we perform simulations to classify the \ac{SHD} dataset~\cite{cramer_etal2020_SHD}. We conduct a parameter sweep over the target sparsity profile $\hat{P}$, which results in different Mosaic architectures with varying connectivity and routing constraints.  For each resulting architecture, we apply the hardware-aware \emph{DeepR} algorithm, incorporating a proxy of the routing constraints into the training process.

\begin{figure}[h]
\centering
\includegraphics[width=0.8\columnwidth]{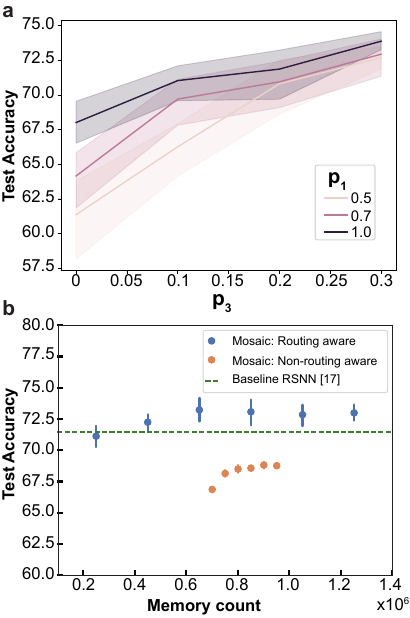}
\caption{Test accuracy of the routing-aware training on the SHD dataset. a)  Test accuracy on SHD for different Mosaic architectures identified by the sparsity of connections for hop 1 and hop 3 (($\hat{P}$)= ($p_1$, $p_3$)), other sparsity set to $0$. The mean and standard deviation of the accuracy is calculated across 30 training runs, for each ($p_1$, $p_3$) pair. b) Test accuracy on SHD as a function of the memory count in the Mosaic architecture, when training using the routing-aware method compared to the non-routing aware one. Both are compared against the vanilla Recurrent Spiking Neural Network (RSNN) results from~\cite{cramer_etal2020_SHD}.}
\label{fig:fig3_results}
\end{figure}

Figure~\ref{fig:fig3_results}a illustrates the accuracy obtained for various $\hat{P}$ profiles. For each $\hat{P}$, we run the simulations for 30 seeds.
We observe that for the (30) networks optimized with a fixed $\hat{P}$, the variance in accuracy ($\sigma_{\text{Acc}}^2$) across seeds is minimal. 
This suggests that $\hat{P}$ is a dominant factor determining the accuracy, regardless of the specific network architecture.

Additionally, Fig.~\ref{fig:fig3_results}b illustrates the relationship between accuracy and the number of memory elements across different $\hat{P}$ profiles. 
Our method achieves comparable accuracy to the \ac{SHD} baseline~\cite{cramer_etal2020_SHD} using a vanilla Recurrent Spiking Neural Network (RSNN). Compared to the non-routing hardware-aware training methods that rely solely on the L1 norm regularizer,~\cite{dalgaty_etal_2024_mosaic}: (i) for the same memory count, our method demonstrates an improved accuracy of 5\%, and (ii) for the same accuracy (68.5\%, maximum achievable by non-routing aware network), it requires an order of magnitude less memory count. 
This approach facilitates effective hardware-software co-design by providing a clear trade-off between memory usage and accuracy for a specific chip configuration.

\section{Conclusions and Discussion}

In this paper, we introduced a hardware-algorithm co-design method to optimize neural networks aware of the underlying hardware architecture. Specifically, we extended the \emph{DeepR} algorithm to not only enforce an upper bound on the number of network parameters, but also ensure its mappability onto a certain memory organization, designed for message passing in a multi-core architecture. This method helps optimizing the architecture-aware neural networks through gradient descent, although the mapping function is often non-differentiable. 

Moreover, we demonstrate that the mapping function can be effectively approximated using a proxy function based on a connection sparsity profile, which constrains the number of  neuron connections based on their spatial distances on the hardware architecture. This approach is validated by the minimal standard deviation in memory resource usage across networks with identical sparsity profile, underscoring the reliability of the proxy for efficient routing-aware network optimization.


Applying this method to the classification of \ac{SHD} spoken digit classification task results in mappable networks onto the hardware architecture, here studied for Mosaic, which utilizes the available memory more optimally than a non-architecture aware trained network, and achieves higher accuracy with the same number of memory elements. 

Other than the better utilization of memory and higher accuracy, there is an interesting observation from the accuracy results of Fig.~\ref{fig:fig3_results}a: 
For the networks with sparsity profile $\hat{P}$, with certain connection sparsity between neurons of certain distance, the standard deviation of the test accuracy across models is minimal. 
This means that the test accuracy might not be a function of the exact connectivity matrix, but rather the connection sparsity profile. 
This might point to the direction that there is a family of network models (and possibly hardware architectures) which perform similarly on a certain task. Whether this can help us design hardware architectures which introduce a good inductive bias for a certain class of problems remains to be further investigated.  



\section{Acknowledgments}
We would like to thank Laura Kriener and Yigit Demirag for helpful comments on the manuscript.

\bibliography{biblio,biblio_mosaic,biblio_algo,biblio_hw}
\bibliographystyle{unsrt}

\end{document}

%% file: acronyms.tex
\acrodef{AI}[AI]{Artificial Intelligence}
\acrodef{ADC}[ADC]{Analog to Digital Converter}
\acrodef{ADEXP}[AdExp-I\&F]{Adaptive-Exponential Integrate and Fire}
\acrodef{AER}[AER]{Address-Event Representation}
\acrodef{AEX}[AEX]{AER EXtension board}
\acrodef{AE}[AE]{Address-Event}
\acrodef{AFM}[AFM]{Atomic Force Microscope}
\acrodef{AGC}[AGC]{Automatic Gain Control}
\acrodef{AMDA}[AMDA]{AER Motherboard with D/A converters}
\acrodef{ANN}[ANN]{Artificial Neural Network}
\acrodef{API}[API]{Application Programming Interface}
\acrodef{ARM}[ARM]{Advanced RISC Machine}
\acrodef{ASIC}[ASIC]{Application Specific Integrated Circuit}
\acrodef{AdExp}[AdExp-IF]{Adaptive Exponential Integrate-and-Fire}
\acrodef{BCM}[BMC]{Bienenstock-Cooper-Munro}
\acrodef{BD}[BD]{Bundled Data}
\acrodef{BEOL}[BEOL]{Back-end of Line}
\acrodef{BG}[BG]{Bias Generator}
\acrodef{BMI}[BMI]{Brain-Machince Interface}
\acrodef{BPTT}[BPTT]{Backpropagation Through Time}
\acrodef{BTB}[BTB]{band-to-band tunnelling}
\acrodef{CAD}[CAD]{Computer Aided Design}
\acrodef{CAM}[CAM]{Content Addressable Memory}
\acrodef{CAVIAR}[CAVIAR]{Convolution AER Vision Architecture for Real-Time}
\acrodef{CA}[CA]{Cortical Automaton}
\acrodef{CCN}[CCN]{Cooperative and Competitive Network}
\acrodef{CDR}[CDR]{Clock-Data Recovery}
\acrodef{CFC}[CFC]{Current to Frequency Converter}
\acrodef{CHP}[CHP]{Communicating Hardware Processes}
\acrodef{CMIM}[CMIM]{Metal-insulator-metal Capacitor}
\acrodef{CML}[CML]{Current Mode Logic}
\acrodef{CMOL}[CMOL]{Hybrid CMOS nanoelectronic circuits}
\acrodef{CMOS}[CMOS]{Complementary Metal-Oxide-Semiconductor}
\acrodef{CNN}[CCN]{Convolutional Neural Network}
\acrodef{COTS}[COTS]{Commercial Off-The-Shelf}
\acrodef{CPG}[CPG]{Central Pattern Generator}
\acrodef{CPLD}[CPLD]{Complex Programmable Logic Device}
\acrodef{CPU}[CPU]{Central Processing Unit}
\acrodef{CSM}[CSM]{Cortical State Machine}
\acrodef{CSP}[CSP]{Constraint Satisfaction Problem}
\acrodef{CV}[CV]{Coefficient of Variation}
\acrodef{DAC}[DAC]{Digital to Analog Converter}
\acrodef{DAS}[DAS]{Dynamic Auditory Sensor}
\acrodef{DAVIS}[DAVIS]{Dynamic and Active Pixel Vision Sensor}
\acrodef{DBN}[DBN]{Deep Belief Network}
\acrodef{DFA}[DFA]{Deterministic Finite Automaton}
\acrodef{DIBL}[DIBL]{drain-induced-barrier-lowering}
\acrodef{DI}[DI]{delay insensitive}
\acrodef{DMA}[DMA]{Direct Memory Access}
\acrodef{DNF}[DNF]{Dynamic Neural Field}
\acrodef{DNN}[DNN]{Deep Neural Network}
\acrodef{DOF}[DOF]{Degrees of Freedom}
\acrodef{DPE}[DPE]{Dynamic Parameter Estimation}
\acrodef{DPI}[DPI]{Differential Pair Integrator}
\acrodef{DRRZ}[DR-RZ]{Dual-Rail Return-to-Zero}
\acrodef{DRAM}[DRAM]{Dynamic Random Access Memory}
\acrodef{DR}[DR]{Dual Rail}
\acrodef{DSP}[DSP]{Digital Signal Processor}
\acrodef{DVS}[DVS]{Dynamic Vision Sensor}
\acrodef{DYNAP}[DYNAP]{Dynamic Neuromorphic Asynchronous Processor}
\acrodef{EBL}[EBL]{Electron Beam Lithography}
\acrodef{EDVAC}[EDVAC]{Electronic Discrete Variable Automatic Computer}
\acrodef{EEG}[EEG]{Electroencephalography}
\acrodef{ECG}[ECG]{Electrocardiography}
\acrodef{EMG}[EMG]{Electromyography}
\acrodef{EIN}[EIN]{Excitatory-Inhibitory Network}
\acrodef{EM}[EM]{Expectation Maximization}
\acrodef{EPSC}[EPSC]{Excitatory Post-Synaptic Current}
\acrodef{EPSP}[EPSP]{Excitatory Post-Synaptic Potential}
\acrodef{ESN}[ESN]{Echo state Network }
\acrodef{EZ}[EZ]{Epileptogenic Zone}
\acrodef{FDSOI}[FDSOI]{Fully-Depleted Silicon on Insulator}
\acrodef{FET}[FET]{Field-Effect Transistor}
\acrodef{FFT}[FFT]{Fast Fourier Transform}
\acrodef{FI}[F-I]{Frequency-Current}
\acrodef{FPGA}[FPGA]{Field Programmable Gate Array}
\acrodef{FR}[FR]{Fast Ripple}
\acrodef{FSA}[FSA]{Finite State Automaton}
\acrodef{FSM}[FSM]{Finite State Machine}
\acrodef{GIDL}[GIDL]{gate-induced-drain-leakage}
\acrodef{GOPS}[GOPS]{Giga-Operations per Second}
\acrodef{GPU}[GPU]{Graphical Processing Unit}
\acrodef{GUI}[GUI]{Graphical User Interface}
\acrodef{HAL}[HAL]{Hardware Abstraction Layer}
\acrodef{HFO}[HFO]{High Frequency Oscillation}
\acrodef{HH}[H\&H]{Hodgkin \& Huxley}
\acrodef{HMM}[HMM]{Hidden Markov Model}
\acrodef{HCS}[HCS]{High-Conductive State}
\acrodef{HRS}[HRS]{High-Resistive State}
\acrodef{HR}[HR]{Human Readable}
\acrodef{HSE}[HSE]{Handshaking Expansion}
\acrodef{HW}[HW]{Hardware}
\acrodef{ICT}[ICT]{Information and Communication Technology}
\acrodef{IC}[IC]{Integrated Circuit}
\acrodef{IEEG}[iEEG]{intracranial electroencephalography}
\acrodef{IF2DWTA}[IF2DWTA]{Integrate \& Fire 2--Dimensional WTA}
\acrodef{IFSLWTA}[IFSLWTA]{Integrate \& Fire Stop Learning WTA}
\acrodef{IF}[I\&F]{Integrate-and-Fire}
\acrodef{IMU}[IMU]{Inertial Measurement Unit}
\acrodef{INCF}[INCF]{International Neuroinformatics Coordinating Facility}
\acrodef{INI}[INI]{Institute of Neuroinformatics}
\acrodef{IO}[I/O]{Input/Output}
\acrodef{IPSC}[IPSC]{Inhibitory Post-Synaptic Current}
\acrodef{IPSP}[IPSP]{Inhibitory Post-Synaptic Potential}
\acrodef{IP}[IP]{Intellectual Property}
\acrodef{ISI}[ISI]{Inter-Spike Interval}
\acrodef{IoT}[IoT]{Internet of Things}
\acrodef{JFLAP}[JFLAP]{Java - Formal Languages and Automata Package}
\acrodef{LEDR}[LEDR]{Level-Encoded Dual-Rail}
\acrodef{LFP}[LFP]{Local Field Potential}
\acrodef{LIF}[LIF]{Leaky Integrate and Fire}
\acrodef{LLC}[LLC]{Low Leakage Cell}
\acrodef{LNA}[LNA]{Low-Noise Amplifier}
\acrodef{LPF}[LPF]{Low Pass Filter}
\acrodef{LCS}[LCS]{Low-Conductive State}
\acrodef{LRS}[LRS]{Low-Resistive State}
\acrodef{LSM}[LSM]{Liquid State Machine}
\acrodef{LTD}[LTD]{Long Term Depression}
\acrodef{LTI}[LTI]{Linear Time-Invariant}
\acrodef{LTP}[LTP]{Long Term Potentiation}
\acrodef{LTU}[LTU]{Linear Threshold Unit}
\acrodef{LUT}[LUT]{Look-Up Table}
\acrodef{LVDS}[LVDS]{Low Voltage Differential Signaling}
\acrodef{MCMC}[MCMC]{Markov-Chain Monte Carlo}
\acrodef{MEMS}[MEMS]{Micro Electro Mechanical System}
\acrodef{MFR}[MFR]{Mean Firing Rate}
\acrodef{MIM}[MIM]{Metal Insulator Metal}
\acrodef{MLP}[MLP]{Multilayer Perceptron}
\acrodef{MOSCAP}[MOSCAP]{Metal Oxide Semiconductor Capacitor}
\acrodef{MOSFET}[MOSFET]{Metal Oxide Semiconductor Field-Effect Transistor}
\acrodef{MOS}[MOS]{Metal Oxide Semiconductor}
\acrodef{MRI}[MRI]{Magnetic Resonance Imaging}
\acrodef{NDFSM}[NDFSM]{Non-deterministic Finite State Machine} 
\acrodef{ND}[ND]{Noise-Driven}
\acrodef{NEF}[NEF]{Neural Engineering Framework}
\acrodef{NHML}[NHML]{Neuromorphic Hardware Mark-up Language}
\acrodef{NIL}[NIL]{Nano-Imprint Lithography}
\acrodef{NMDA}[NMDA]{N-Methyl-D-Aspartate}
\acrodef{NME}[NE]{Neuromorphic Engineering}
\acrodef{NN}[NN]{Neural Network}
\acrodef{NRZ}[NRZ]{Non-Return-to-Zero}
\acrodef{NSM}[NSM]{Neural State Machine}
\acrodef{OR}[OR]{Operating Room}
\acrodef{OTA}[OTA]{Operational Transconductance Amplifier}
\acrodef{PCB}[PCB]{Printed Circuit Board}
\acrodef{PCHB}[PCHB]{Pre-Charge Half-Buffer}
\acrodef{PCM}[PCM]{Phase Change Memory}
\acrodef{PE}[PE]{Processing Element}
\acrodef{PFA}[PFA]{Probabilistic Finite Automaton}
\acrodef{PFC}[PFC]{prefrontal cortex}
\acrodef{PFM}[PFM]{Pulse Frequency Modulation}
\acrodef{PR}[PR]{Production Rule}
\acrodef{PSC}[PSC]{Post-Synaptic Current}
\acrodef{PSP}[PSP]{Post-Synaptic Potential}
\acrodef{PSTH}[PSTH]{Peri-Stimulus Time Histogram}
\acrodef{QDI}[QDI]{Quasi Delay Insensitive}
\acrodef{RAM}[RAM]{Random Access Memory}
\acrodef{RDF}[RDF]{random dopant fluctuation}
\acrodef{RELU}[ReLu]{Rectified Linear Unit}
\acrodef{RLS}[RLS]{Recursive Least-Squares}
\acrodef{RMSE}[RMSE]{Root Mean Squared-Error}
\acrodef{RMS}[RMS]{Root Mean Squared}
\acrodef{RNN}[RNN]{Recurrent Neural Networks}
\acrodef{RSNN}[RSNN]{Recurrent Spiking Neural Network}
\acrodef{ROLLS}[ROLLS]{Reconfigurable On-Line Learning Spiking}
\acrodef{RRAM}[RRAM]{Resistive Random Access Memory}
\acrodef{R}[R]{Ripples}
\acrodef{SAC}[SAC]{Selective Attention Chip}
\acrodef{SAT}[SAT]{Boolean Satisfiability Problem}
\acrodef{SCX}[SCX]{Silicon CorteX}
\acrodef{SD}[SD]{Signal-Driven}
\acrodef{SEM}[SEM]{Spike-based Expectation Maximization}
\acrodef{SLAM}[SLAM]{Simultaneous Localization and Mapping}
\acrodef{SNN}[SNN]{Spiking Neural Network}
\acrodef{SNR}[SNR]{Signal to Noise Ratio}
\acrodef{SOC}[SOC]{System-On-Chip}
\acrodef{SOI}[SOI]{Silicon on Insulator}
\acrodef{SP}[SP]{Separation Property}
\acrodef{SHD}[SHD]{Spiking Heidelberg Digit}
\acrodef{SSC}[SSC]{Spiking Speech Command}
\acrodef{SRAM}[SRAM]{Static Random Access Memory}
\acrodef{SRNN}[SRNN]{Spiking Recurrent Neural Network}
\acrodef{STDP}[STDP]{Spike-Timing Dependent Plasticity}
\acrodef{STD}[STD]{Short-Term Depression}
\acrodef{STP}[STP]{Short-Term Plasticity}
\acrodef{STT-MRAM}[STT-MRAM]{Spin-Transfer Torque Magnetic Random Access Memory}
\acrodef{STT}[STT]{Spin-Transfer Torque}
\acrodef{SW}[SW]{Software}
\acrodef{TCAM}[TCAM]{Ternary Content-Addressable Memory}
\acrodef{TFT}[TFT]{Thin Film Transistor}
\acrodef{TPU}[TPU]{Tensor Processing Unit}
\acrodef{USB}[USB]{Universal Serial Bus}
\acrodef{VHDL}[VHDL]{VHSIC Hardware Description Language}
\acrodef{VLSI}[VLSI]{Very Large Scale Integration}
\acrodef{VOR}[VOR]{Vestibulo-Ocular Reflex}
\acrodef{WCST}[WCST]{Wisconsin Card Sorting Test}
\acrodef{WTA}[WTA]{Winner-Take-All}
\acrodef{XML}[XML]{eXtensible Mark-up Language}
\acrodef{CTXCTL}[CTXCTL]{CortexControl}
\acrodef{divmod3}[DIVMOD3]{divisibility of a number by three}
\acrodef{hWTA}[hWTA]{hard Winner-Take-All}
\acrodef{sWTA}[sWTA]{soft Winner-Take-All}
\acrodef{APMOM}[APMOM]{Alternate Polarity Metal On Metal}
\acrodef{SRNN}[SRNN]{Spiking Recurrent Neural Networks}
\acrodef{fMRI}[fMRI]{functional Magnetic Resonance Imaging}
\acrodef{RL}[RL]{Reinforcement Learning}
\acrodef{ES}[ES]{Evolutionary Strategies}

\acrodef{NT}[NT]{Neuron Tile}
\acrodef{RT}[RT]{Routing Tile}